\documentclass[letterpaper,aps,prl,twocolumn,amsmath,showpacs,amssymb]{revtex4}

\usepackage{epsfig}
\usepackage{graphicx}
\usepackage{dcolumn}
\usepackage{bm}
\usepackage{bbm}
\usepackage{wasysym}
\usepackage{marvosym}
\usepackage{amsmath,amssymb}

\newlength{\textwidthm}

\usepackage{ulem}
\usepackage{color}
\begin{document}

\title{Low density ferromagnetism in biased bilayer graphene}

\author{Eduardo~V.~Castro,$^1$ N.~M.~R.~Peres,$^2$ T.~Stauber,$^{2,3}$ and
N.~A.~P.~Silva$^2$}

\affiliation{$^1$CFP and Departamento de F\'{\i}sica, 
  Faculdade de Ci\^encias Universidade do Porto, P-4169-007 Porto, Portugal}

\affiliation{$^2$Centro de F\'{\i}sica  e  Departamento de
F\'{\i}sica, Universidade do Minho, P-4710-057, Braga, Portugal}

\affiliation{$^3$Instituto de Ciencia de Materiales de
Madrid. CSIC. Cantoblanco. E-28049 Madrid, Spain}

\date{\today}

\begin{abstract}
We compute the phase diagram of a biased graphene bilayer.
The existence of a ferromagnetic phase is discussed with
respect both to carrier density and temperature.
We find that the ferromagnetic transition is
first order, lowering the value of $U$ relatively to the usual Stoner
criterion.
We show that in the ferromagnetic phase
the two planes have unequal magnetization and that the electronic
density is hole like in one plane and electron like in the other.
\end{abstract}


\maketitle

{\it Introduction.---}Graphene, a two-dimensional hexagonal lattice of 
carbon atoms, has attracted 
considerable attention due to its unusual electronic properties, 
characterized by massless Dirac 
fermions \cite{NGP+rmp07,Katsnelson07}. It was first produced 
by micromechanical cleavage of graphite
and its hallmark  is the half integer quantum Hall effect \cite{Nov07}.

In addition to graphene, few-layer graphene can also be produced. Of particular
interest to us is {\it bilayer graphene} (BLG), where 
two carbon planes lay on top of each other
 according to  $AB$-Bernal stacking. 
In BLG it is possible to 
have the two planes at different
electrostatic potentials. 
As a consequence, a gap opens 
at the Dirac point and the low energy band acquires a Mexican hat 
dispersion \cite{Guinea}.  This system is called a biased BLG, 
and provides the first semiconductor with a gap that can be
tuned externally  \cite{Ohta,Eduardo,Oostinga}.
Due to the Mexican hat dispersion the density of states (DOS) 
close
to the gap diverges as the square root of the energy. The possibility
of having an arbitrary large DOS at the Fermi energy
poses the question whether this system can be unstable toward a
ferromagnetic ground state -- a question we want to address in this Letter.  
From the point of view of the
exchange instability, BLG was found to be always unstable toward a 
ferromagnetic ground state for low enough densities \cite{Nilsson06,Sta07}.

The question of magnetism in carbon based
systems has already a long history.  Even before the discovery of
graphene, graphite has attracted a
broad interest due to the observation of anomalous properties, such as
magnetism and insulating behavior in the direction perpendicular to
the planes \cite{Eetal02,KEK02,Ketal03b,Kopelevich03,Ohldag07}. The research of
$s-p$ based magnetism \cite{Rode04,Turek91,Srdanov98} was especially
motivated by the technological use of nanosized particles of graphite,
which show interesting features depending on their shape, edges, and
applied field of pressure \cite{Enoki05}.
Microscopic theoretical models of bulk carbon magnetism include
nitrogen-carbon compositions where ferromagnetic ordering of spins
could exist in $\pi$ delocalized systems due to a lone electron pair
on a trivalent element \cite{Ovch78} or intermediate graphite-diamond
structures where the alternating $sp^2$ and $sp^3$ carbon atoms play
the role of different valence elements \cite{Ovch91}. More general
models focus on the interplay between disorder and
interaction \cite{Sta05,Voz05}. Further, midgap states due to zigzag
edges play a predominant role in the formation of magnetic moments
\cite{Fujita,Pisani07} which support flat-band
ferromagnetism \cite{Mielke91,Tasaki98,Kus03}. Magnetism is also found
in fullerene based metal-free systems \cite{Chan04}. For a recent
overview on metal-free carbon based magnetism see
Ref.~[\onlinecite{Makarova}].

{\it Model and mean field treatment.---}Due to the electrostatically 
invoked band-gap, there is a large
DOS for low carrier density and thus effective screening of 
the Coulomb interaction. Coulomb interaction shall thus be treated 
using a Hubbard on-site interaction.

The Hamiltonian of a biased BLG Hubbard model is the sum of two
pieces $H=H_{TB}+H_U$, where $H_{TB}$ is the tight-binding part 
and $H_U$ is the Coulomb on-site interaction part.
The term $H_{TB}$ is a sum of four terms: 
the tight-binding Hamiltonian of each plane, the hopping term between 
planes, and the applied electrostatic bias.
We therefore have
$
H_{TB} = \sum_{\iota = 1}^2 H_{TB,\iota}+H_\perp+H_V\,,
$
with
$
H_{TB,\iota} =-t\sum_{\bm r,\sigma}a^\dag_{\iota\sigma}(\bm r)
[b_{\iota\sigma}(\bm r)
+
b_{\iota\sigma}(\bm r-\bm a_1)
+
b_{\iota\sigma}(\bm r-\bm a_2)]
+
h.c.$, 
$
H_\perp = -t_\perp\sum_{\bm r,\sigma}
[a^\dag_{1\sigma}(\bm r)b_{2\sigma}(\bm r)+
h.c.
]$,
and
$
H_V = \frac V 2\sum_{\bm r, x,\sigma}
[n_{x1\sigma}(\bm r)-
n_{x2\sigma}(\bm r)]\,.
$
The term $H_U$ is given by
$
H_U=U\sum_{\bm r, x}[n_{x1\uparrow}(\bm r)n_{x1\downarrow}(\bm r)+
n_{x2\uparrow}(\bm r)n_{x2\downarrow}(\bm r)]$.
We used 
$n_{x\iota\sigma}(\bm r)=x^\dag_{\iota\sigma}(\bm r)x_{\iota\sigma}(\bm r)$, 
$x=a(b)$,
as the number
operator at position $\bm r$ and sublattice $A\iota$ ($B\iota$) of 
layer $\iota=1,2$, for spin $\sigma=\uparrow,\downarrow$;
$\mathbf{a}_{1}=a(1,0)$ and $\mathbf{a}_{2}=a(1,-\sqrt{3})/2$ are
the basis vectors and $a\approx2.46\,\textrm{\AA}$ the lattice
constant.
Unless stated otherwise, we use 
$t=2.7$~{\ttfamily eV}, $t_\perp=0.2t$, and
$V=0.05$~{\ttfamily eV} \cite{foot0}.

The problem defined by $H_{TB}+H_U$ 
cannot be solved exactly.
We adopt a mean field approach, recently applied to describe
magnetic properties of graphene nanoislands \cite{RP07}.
Since the two
planes of the BLG are at different electrostatic potentials, 
we expect an asymmetry between layers for the charge density $n$
and the magnetization $m=n_\uparrow-n_\downarrow$ (per unit cell).
Accordingly, we propose the following 
broken symmetry ground state,
which also defines the mean field parameters:
$
\langle n_{x 1\sigma}(\bm r)\rangle =\frac {n+\Delta n} 8 
+ \sigma \frac {m+\Delta m} 8
$
and
$
\langle n_{x 2\sigma}(\bm r)\rangle =\frac {n-\Delta n} 8 
+ \sigma \frac {m-\Delta m} 8\,,
$
where $\Delta n$ and $\Delta m$ represent the charge
density and  the spin polarization difference
between the two layers, respectively \cite{foot1}.
This leads to an effective bias
$V_\sigma=V+U\Delta n/4-\sigma U \Delta m/4$.

If one assumes the ferromagnetic transition to be second order,
with $m = 0$ and $\Delta m = 0$ at the transition,
 we are lead to a $U-$critical $U_c$ given by,
\begin{equation}
U_c=1/\rho_b(\tilde\mu,U_c)\,,
\label{Eq_stoner}
\end{equation}
where  $\rho_b(\tilde\mu,U_c)$ is the DOS per spin per 
lattice point and $\tilde \mu = \mu - nU_c/8$, with 
 chemical potential $\mu$. 
Although Eq.~(\ref{Eq_stoner}) looks like
the usual Stoner criterion,   
the effective bias $V_\sigma$  depends on $U$ due to $\Delta n$.
This makes Eq.~(\ref{Eq_stoner}) non-linear,
and $U_c$ has to be found numerically in a self-consistent way.

\begin{figure}[t]
\begin{center}
\includegraphics*[angle=0,width=0.9\columnwidth]{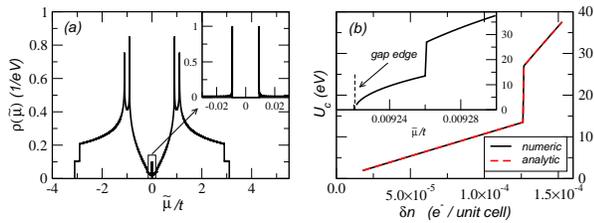}
\end{center}
\caption{(Color online) (a)~Bilayer graphene DOS for $U=0$. 
  Inset: Zoom near the gap region. 
  (b)~$U_c$ vs $\delta n$ in the low
  doping regime. Inset: The same as a function of $\tilde \mu$.
}
\label{Fig_DOS}
\end{figure}

{\it Simple results.---}We start with the zero 
temperature ($T=0$) phase diagram
in the plane $U$~vs~$\delta n$, where $\delta n$
is 
the doping relatively to the half filled case. An approximate
analytic treatment is possible in this limit, which is
used to check our numerical results. 

In Fig.~\ref{Fig_DOS}~(a) we represent the DOS
of a biased BLG with $U=0$.
As seen in the inset, 
the DOS diverges at the edges of the gap. As a consequence,
the closer the chemical potential to the gap edges, the
lower  the critical $U_c$ value. The low doping 
$U_c$ value -- 
given by Eq.~(\ref{Eq_stoner}) in the limit 
$U\Delta n \ll V$ -- is shown in Fig.~\ref{Fig_DOS}~(b),
both as a function of $\delta n$ and $\tilde \mu$ (inset). The lowest
represented value of $U_c$ is about $U_c \simeq 2.7$~{\ttfamily eV}
to which corresponds 
$\delta n \simeq 2.5\times10^{-5}$ electrons per unit cell. 
The step like discontinuity shown in panel~(b) for $U_c$ 
occurs when the Fermi energy equals $V/2$, signaling
the top of the Mexican hat dispersion relation.

It is clear from  Fig.~\ref{Fig_DOS}~(b)
that in the low doping limit $U_c$ is a linear function of $\delta n$.
To understand this behavior, first we note that for very low doping
the DOS  close to the gap edges behaves as
$
\rho_b (\tilde \mu) \propto (|\tilde \mu| - \Delta_g/2)^{-1/2}\,,
$
where $\Delta_g$ is the size of the gap.
Using this approximate expression to compute the doping,
$\delta n \propto \textrm{sign}(\tilde \mu) \times 
\int_{\Delta_g/2}^{|\tilde \mu|} \textrm{d}x~\rho_b(x)$, we immediately get 
$\delta n \propto \textrm{sign}(\tilde \mu)/\rho_b (\tilde \mu)$ 
and thus $U_c \propto |\delta n|$. 
In Fig.~\ref{Fig_DOS}~(b) both the numerical result
of Eq.~(\ref{Eq_stoner}) and the approximated analytical result just derived
are shown. The agreement is excellent.

\begin{figure}[t]
\begin{center}
\includegraphics*[angle=0,width=0.98\columnwidth]{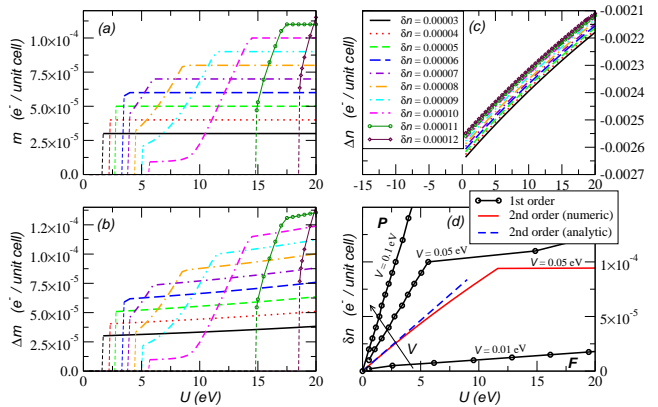}
\end{center}
\caption{(Color online) 
Panels~(a), (b), and~(c) show the $T=0$ solution 
for~$m$, $\Delta m$, and $\Delta n$, respectively. 
Panel~(d) shows the $U$ vs $\delta n$
phase diagram at  $T = 0$: symbols are
inferred from panel~(a) and signal a {\it first-order} transition;
lines stand for the {\it second-order} one given 
by Eq.~(\ref{Eq_stoner}).
Labels: $P$-paramagnetic, $F$-ferromagnetic.
}
\label{Fig_PD_n}
\end{figure}

{\it Self-consistent solution.---}In order to obtain the $T=0$ phase diagram
 of the biased BLG, we study how $m$, $\Delta m$,
and $\Delta n$ depend on the interaction $U$, for given
values of the electronic doping $\delta n$. 

In Fig.~\ref{Fig_PD_n}~(a)
 it is shown how $m$ depends
on $U$ for different values of $\delta n$. The chosen values of
$\delta n$ correspond to the chemical potential being located at
the divergence of the low energy DOS, which explains the
smaller critical $U_{c}$ value for smaller $\delta n$.
 It is interesting to note that the 
saturation values of the magnetization
correspond to full polarization of the doping charge density
with $m=\delta n$, also found within a one-band model \cite{Sta07} .
In Fig.~\ref{Fig_PD_n}~(b) we plot
the $\Delta m$ vs $U$.
Interestingly, the value of $\Delta m$
vanishes at the same $U_{c}$ as $m$. For finite values of $m$ we
have $\Delta m>m$, which means that the magnetization of the two
layers is opposite and unequal.
 In Fig.~\ref{Fig_PD_n}~(c)
we show $\Delta n$ vs $U$.
 It is clear
that $|\delta n|<|\Delta n|$, which implies that the density of charge
carriers is above the Dirac point in one plane and below it in the
other plane. This means that the charge carriers are electron like
in one plane and hole like in the other. As $U$ is
increased $\Delta n$ is suppressed in order
to reduce the system Coulomb energy.

In Fig.~\ref{Fig_PD_n}~(d) we show the 
$T=0$ phase diagram  in the
$U$ vs $\delta n$ plane. 
Here we concentrate on the $V = 0.05$~{\ttfamily eV} case.
Symbols are inferred from the magnetization in panel~(a). 
They signal a \textit{first-order} transition
when $m$ increases from zero to a finite value {[}see panel~(a)].
The full (red) line is the numerical self-consistent result of 
Eq.~(\ref{Eq_stoner}),
and the dashed (blue) line is the approximate analytic result 
described above. The discrepancy between lines and symbols
has a clear meaning. In order to obtain Eq.~(\ref{Eq_stoner})
we assumed that a \textit{second-order} 
transition would take place.
This is not the case, and the system undergoes a first-order 
transition for smaller $U$ values. There are clearly two different regimes:
 one for $\delta n\lesssim 10^{-4}$,
where the dependence of $\delta n$ on $U_{c}$ is linear, and another
 for $\delta n > 10^{-4}$, where a plateau like behavior
develops. This plateau has the same physical origin as the step like
discontinuity we have seen in 
Fig.~\ref{Fig_DOS}~(b).
In the limit $\delta n\rightarrow0$
we have not only $U_{c}\rightarrow0$, but also
$m\rightarrow0$ and $\Delta m\rightarrow0$ [see panels~(a) and~(b) 
of Fig.~\ref{Fig_PD_n}], implying
a \emph{paramagnetic} ground state for the undoped
biased BLG. 

Figure~\ref{Fig_PD_n}~(d) shows also 
the effect of $V$ on the
$T=0$ phase diagram (the effect of $t_\perp$
being similar). Raising either $V$
or $t_{\perp}$ leads to a decrease of the critical-$U$
needed to establish the ferromagnetic phase for a given $\delta n$.
The order of the transition, however, remains \emph{first-order}.
We have observed that decreasing $t_{\perp}$ leads to a decrease
in $\Delta m$, and below some $t_{\perp}$ we can have $\Delta m<m$.
A similar effect has been seen when $V$ is increased. It should
be noted, however, that $m$ and $\Delta m$ are $U$-dependent,  meaning that,
depending on  $V$ and $t_{\perp}$, we
can go from $\Delta m<m$ to $\Delta m>m$ just by increasing $U$. Irrespective
of $V$ and $t_{\perp}$ we have always observed $|\delta n|<|\Delta n|$:
electron like carriers in one plane and hole like in the other. 

\begin{figure}[t]
\begin{centering}
\includegraphics[width=0.7\columnwidth]{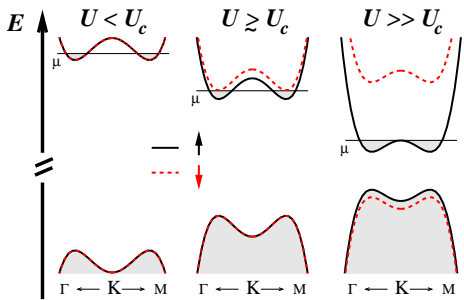}
\par\end{centering}
\caption{\label{fig:bs}
(Color online) Hartree-Fock bands for $\uparrow$ (full lines) and
  $\downarrow$ (dashed lines) spin polarizations.}
\end{figure}

{\it Understanding the asymmetry between planes.---}The asymmetry between 
planes regarding both charge and spin polarization
densities can be understood based on the Hartree-Fock bands shown
in Fig.~\ref{fig:bs}. 
Additionally, we note that in the biased BLG the weight of
the wave functions in each layer for near-gap states is strongly dependent
on their valence band or conduction band character \cite{McC06,Eduardo,MSB+06}.
Valence band states near the gap are mostly localized
on layer~2, due to the lower electrostatic potential $-V/2$. 
On the other hand, near-gap conduction band states
have their highest amplitude on layer~1, due to the higher electrostatic
potential $+V/2$.

The case $U<U_{c}$ shown in Fig.~\ref{fig:bs} (left) stands for
the paramagnetic phase. The values $m=0$ and $\Delta m=0$ 
are an immediate consequence of the degeneracy of $\uparrow$
and $\downarrow$ spin polarized bands. The presence of a finite gap,
however, leads to the abovementioned asymmetry between near-gap valence
and conduction states. As a consequence, a half-filled BLG would
have $n_{2}=(4+\Delta n)/2$~\texttt{e$^{-}$/unit cell} 
on layer~2 (electron like  carriers) and 
$n_{1}=(4-\Delta n)/2$~\texttt{e$^{-}$/unit cell}
 on layer~1 (hole like  carriers), with $\Delta n\neq0$.
Even though the system  is not at half-filling, as long
as $|\delta n|<|\Delta n|$ the carriers on layers~1 and~2 will
still be hole and electron like, respectively. 

Let us now consider the case $U\gtrsim U_{c}$ shown in Fig.~\ref{fig:bs}
(center). The degeneracy lifting of spin polarized bands gives rise
to a finite magnetization, $m\neq0$. Interestingly enough, the degeneracy
lifting is only appreciable for conduction bands, as long as $U$
is not much higher than $U_{c}$. This explains why 
we have $m\approx\Delta m$, 
as shown in panels~(a) and~(b)
of Fig.~\ref{Fig_PD_n} -- as only conduction bands are contributing
to $\Delta m$, the spin polarization density is almost completely
localized in layer~1, where $m_{1}=(m+\Delta m)/2\approx m$, while
the spin polarization in layer~2 is negligible, 
$m_{2}=(m-\Delta m)/2\approx0$. 

It is only when $U\gg U_{c}$ that valence bands become non-degenerate,
as seen in Fig.~\ref{fig:bs} (right). This implies that near-gap
valence states with $\uparrow$ and $\downarrow$ spin polarization have
different amplitudes in layer~2. As the valence band for $\downarrow$
spin polarization has a lower energy the near-gap valence states with
spin $\downarrow$ have higher amplitude in layer~2 than their spin
$\uparrow$ counterparts. Consequently, the magnetization in layer~2
is effectively opposite to that in layer~1, i.e., $\Delta m>m$,
as can be observed in panels~(a) and~(b) of Fig.~\ref{Fig_PD_n}.

We note that the cases $U\gtrsim U_{c}$ and $U\gg U_{c}$
are parameter dependent. The valence
bands can show an appreciable degeneracy lifting already for $U\gtrsim U_{c}$,
especially for small values of the $t_{\perp}$ parameter.
In this case the magnetization of the two layers is no longer opposite,
with $\Delta m<m$. This can be understood as due to the fact that
as $t_{\perp}$ is decreased the weight of near-gap wave functions
becomes more evenly distributed between layers, leading not only to
a decrease in $\Delta n$ but also in $\Delta m$.

%
\begin{figure}[t]
\begin{centering}
\includegraphics*[width=0.98\columnwidth]{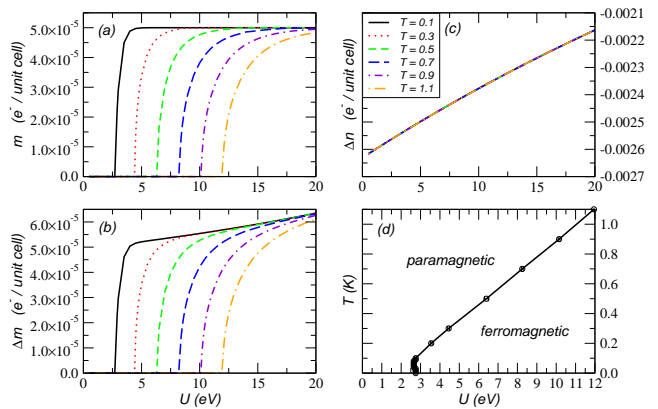} 
\par\end{centering}
\caption{(Color online) Panels~(a), (b), and~(c) show the 
finite $T$
solution for~$m$, $\Delta m$, and $\Delta n$,
respectively, with $T$ measured in~\texttt{K}. 
Panel~(d) shows the $U$ vs $T$ phase diagram. 
}
\label{Fig_PD_T} 
\end{figure}

{\it Finite temperature.---}Now we describe the phase diagram of the 
biased BLG in the 
$T$ vs  $U$ plane. This is done in Fig.~\ref{Fig_PD_T} for 
$\delta n=5\times10^{-5}$~\texttt{e$^{-}$/unit cell}.
For $T=0-1.1$~\texttt{K}
we studied the dependence of $m$, $\Delta m$ and $\Delta n$ on
the  interaction $U$. First we note that the minimum
critical-$U$ is not realized 
at $T=0$. There
is a reentrant behavior which is signaled by the smallest $U_{c}$
for $T=0.06\pm0.02$\texttt{~K}. For temperatures above $T\approx0.1$\texttt{~K}
we have larger $U_{c}$ values for the larger temperatures, as can
be seen in panel~(a). The same is true for $\Delta m$ in panel~(b).
As in the case of Fig.~\ref{Fig_PD_n}, the value of $\Delta m$,
at a given $T$ and $U$, is larger than $m$. 
Also the value of $\Delta n$, shown in panel~(c), is larger than $\delta n$.
Therefore we have the two planes presenting opposite magnetization
and the charge carriers being hole like in one graphene plane and
electron like in the other. 
In panel~(d) of Fig.~\ref{Fig_PD_T}
we present the phase diagram in the $U$ vs $T$. Except at very
low temperatures, there is a linear dependence of $U_{c}$ on $T$.
It is clear that at low temperatures, $T\simeq$ 0.2\texttt{~K},
the value of $U_{c}$ is smaller than the estimated values of $U$
for carbon compounds \cite{Parr50,Baeriswyl86}.

{\it Disorder.---}Crucial prerequisite in order to find ferromagnetism
is a high DOS at the Fermi energy.  The presence of
disorder will certainly cause a smoothing of the singularity in the
DOS and the band gap renormalization, and can even lead
to the closing of the gap.  We note, however, that for small values of
the disorder strength the DOS still shows an enhanced
behavior at the band gap edges \cite{Nilsson07}. The 
strong suppression of electrical noise in BLG \cite{LA08}
further suggests that in
addition to a high crystal quality -- leading to remarkably high 
mobilities~\cite{Morozov08} -- an effective screening of random
potentials is at work.
Disorder should thus not be a limiting factor
in the predicted low density ferromagnetic state, as long as
standard high quality BLG samples are concerned.

Let us also comment on the next-nearest interlayer-coupling
$\gamma_3$, which in the unbiased case breaks the spectrum into four
pockets for low densities \cite{McCann06}. In the biased case,
$\gamma_3$ still breaks the cylindrical symmetry, leading to the
trigonal distortion of the bands, but the divergence in the density of
states at the edges of the band gap is
preserved \cite{Nilsson07}. Therefore, the addition of $\gamma_3$ to
the model does not qualitatively change our result.

{\it Conclusion.---}We have found that in the ferromagnetic phase 
the two layers in
 general have opposite magnetization and that the electronic density
 is hole like in one plane and electron like in the other. We have
 also found that at zero temperature, where the transition can be
 driven by doping, the phase transition between paramagnetic and
 ferromagnetic phases is \emph{first-order}.

EVC, NMRP and TS acknowledge the financial support from POCI 2010
via project PTDC/FIS/64404/2006, the ESF Science Program INSTANS.
This work has also been
supported by MEC (Spain) through Grant No. FIS2004-06490-C03-00, by
the European Union, through contract 12881 (NEST), and the Juan de la
Cierva Program (MEC, Spain).


\begin{thebibliography}{99}
\bibitem{foot0}
We assume that $V$ and the charge density are independent
and can be externally controlled \cite{Oostinga}.
\bibitem{foot1}
Assuming equal spin densities in sublattices $A$ and $B$ of the
same layer is a reasonable approximation for $t_\perp \ll t$.
\end{thebibliography}
\end{document}